\documentclass[11pt]{article}

\usepackage[margin=1.1in]{geometry}
\usepackage{amsmath,amssymb,amsthm}
\usepackage{mathtools}
\usepackage{booktabs}
\usepackage{graphicx}
\usepackage{enumitem}
\usepackage[round]{natbib}
\usepackage[colorlinks=true,linkcolor=blue,citecolor=blue,urlcolor=blue]{hyperref}
\usepackage{microtype}

\newtheorem{theorem}{Theorem}[section]
\newtheorem{proposition}[theorem]{Proposition}
\newtheorem{definition}[theorem]{Definition}

\newtheorem{assumption}{Assumption}

\newtheorem{hypothesis}{Hypothesis}

\theoremstyle{remark}
\newtheorem{remark}{Remark}[section]

\newcommand{\E}{\mathbb{E}}
\newcommand{\Prob}{\mathbb{P}}

\newcommand{\F}{\mathcal{F}}

\newcommand{\abn}{\mathrm{abn}}

\title{Extreme Value Alpha and Crash Risk:\\
Separating Structural Tails from Lottery Tails with LLM-Extracted Disclosure Networks}

\author{Lin Zhang\thanks{VinePeak Research; \texttt{linzhang1126@gmail.com} (corresponding author).}
\and
Fan Yang\thanks{VinePeak Research; \texttt{fyang@vinepeak.com}.}
}

\date{August 2026}

\begin{document}

\maketitle

\begin{abstract}
\noindent
A heavy upper tail in a stock's return distribution is ambiguous: it can be a \emph{lottery tail}---transient jump risk that investors systematically overpay for, earning the well-documented MAX discount---or a \emph{structural tail}, the statistical shadow of an economic reconfiguration that precedes historical extreme winners. Returns alone cannot separate the two, so tail heat alone is not an alpha signal. We propose the firm's \emph{disclosure-measured economic network} as the discriminator: a directed, span-grounded graph extracted from 10-K filings by an auditable LLM pipeline. Its vintage-to-vintage rewiring decomposes exactly into edge birth, death, and drift, producing the paper's central sign pattern: \emph{tail heat with network death is the crash side; tail heat with an intact or forming network is where structural tails---and historical winners---live.}

Pilot evidence from 24 technology firms (2014--2025) already supports the crash side. The interaction of upper-tail heat with death mass predicts negative forward abnormal returns (monthly $t\approx-2.9$; collapsed to 200 firm-vintages, $t\approx-3.9$; wild-cluster bootstrap $p=0.04$; robust to two-way clustering and to MAX, momentum, and volatility controls), and the same configuration preceded NVIDIA's 2018 and 2022 drawdowns. This death-side signal is immediately useful as a risk-monitoring danger flag. The alpha side---the claim that the market misprices structural tails as lotteries until the filings' configuration becomes common knowledge, leaving a category-correction premium---is directionally positive but not significant in the pilot and awaits the pre-specified confirmatory test.

A pre-registered replication on 50 randomly sampled S\&P~500 firms \emph{failed}, and the failure defines the mechanism's boundary: outside coherent ecosystems the disclosure graph nearly vanishes (83\% of firm-vintages carry zero death mass), so the discriminator is defined only where firms densely document counterparties. This scope condition bounds all claims.

The confirmatory design is fully pre-specified---fixed walk-forward windows, a gatekept primary pair, archived power simulations for both endpoints, selection-corrected benchmarks, positive-only winner labels, and a frozen ecosystem-coherent universe rule with a density gate; a completed recall audit has moved extraction to a multi-agent protocol, with the cutoff-matched extractor audit still pending. The design activates only if its density gate passes. If confirmed, it would turn tail heat from a lottery screen into a conditional signal that separates crash risk from structural winners---and disclosure text into the instrument that tells them apart.\\[6pt]
\textbf{Keywords:} extreme value theory; lottery stocks; MAX anomaly; large language models; corporate disclosures; economic networks; tail risk.\\
\textbf{JEL:} C46, C58, G11, G12, G14.
\end{abstract}

\section{Introduction}\label{sec:intro}

\subsection{One question with two sides}

This paper asks one question with two sides: \emph{which heavy tails carry alpha, and which carry crash risk?}

On the alpha side, two literatures give contradictory instincts. The lottery-stock literature says no heavy tail carries alpha: stocks with extreme recent upside---high maximum daily returns---systematically \emph{underperform}, because investors with skewness preference overpay for jump exposure \citep{bali2011max,barberis2008lotteries}. The case-study record says some do: nearly every historical extreme winner (NVIDIA is this paper's running example) exhibited a heating upper tail before and during its run, and an investor who treated that heat as a lottery symptom missed the defining returns of the decade. On the crash side, the instincts are less contradictory but no less conditional: crash risk is forecastable \citep{chen2001crash}, and it concentrates where disclosure obscures what is happening inside the firm \citep{hutton2009opacity}---yet a hot upper tail is, on its face, the \emph{opposite} of a crash signature, which is precisely why the crashes that follow hot tails (NVIDIA 2018, 2022) blindside screens built on returns alone. Both sides point to the same resolution: heavy tails come in (at least) two kinds, and the same observable---tail heat---sits on top of both.

We give the two kinds names and an observable discriminator:

\begin{itemize}[leftmargin=1.6em]
\item A \textbf{lottery tail} is generated by transient, idiosyncratic jump risk---news shocks, short squeezes, retail attention---with no persistent economic reconfiguration behind it. Skewness-preferring investors overpay for it; its conditional alpha is negative (the MAX discount).
\item A \textbf{structural tail} is the statistical shadow of a regime shift in the firm's economic configuration: new dependencies forming, counterparty structure recomposing, a platform position emerging. Its extreme returns are draws from a persistent process that is still repricing; its conditional alpha is non-negative and, we will argue, positive while the configuration diffuses.
\end{itemize}

The discriminator we propose is the firm's \emph{disclosure-measured economic network}. Large language models can extract, from 10-K filings, a directed, weighted, point-in-time graph of documented economic relationships---supplier, customer, licensing, cloud, and platform dependencies---with every edge grounded in quoted disclosure text \citep{yang2026lem}. Vintage-to-vintage rewiring of this graph decomposes \emph{exactly} into edge-birth mass $B$, edge-death mass $D$, and continuing drift $C$. The central empirical claim of the paper is a sign pattern:

\begin{quote}
\emph{Tail heat $\times$ network death is the crash side; tail heat with an intact or forming network is where structural tails---and historical winners---live.}
\end{quote}

\subsection{What the pilot data already show}

Three descriptive facts from our 24-firm technology panel (2014--2025) motivate everything that follows. First, \emph{unconditional tail heat is not an alpha signal}: forward six-month abnormal returns are flat across quintiles of the rolling upper-tail index, and raw quintiles of tail asymmetry (right tail heavier than left) slope toward \emph{under}performance---consistent with the lottery discount operating even among mega-cap technology names, though attenuated with controls (Section~\ref{sec:pilot-h1}). Second, \emph{the case study partially survives in event time}: NVIDIA's upper-tail index first crossed the panel's 80th percentile in February 2016, at the onset of and ahead of the bulk of its 2016--18 run; it collapsed through 2019--22, and rose again through the 2023--25 run, crossing the threshold only in May 2024---a timing lead for the first run, a lag for the second. Third---the paper's pivotal finding---\emph{the network separates the two kinds}: interacting tail heat with the rewiring components, the negative interaction loads entirely on death mass ($t\approx-2.9$ monthly, $t\approx-3.9$ collapsed to firm-vintages, wild-cluster $p=0.04$, with lottery, momentum, and volatility controls), while tail heat interacted with birth mass is statistically indistinguishable from zero. The two crashes in NVIDIA's decade (2018, 2022) were both preceded by death-dominated rewiring under a still-hot tail; both runs began from birth-dominated rewiring.

\subsection{Where the alpha---and the crash risk---come from}

Both sides of the sign pattern flow from the same friction: the market observes tail heat directly (it is in every screen), while the configuration behind the tail is documented only in disclosure text, where it diffuses slowly \citep{cohen2020lazy}.

On the alpha side, if the pattern survives confirmation, the economics is a \emph{category-correction premium}. The lottery discount is applied broadly to observed heat; a structural tail therefore trades at a discount it does not deserve, and the alpha accrues as the configuration becomes common knowledge---through earnings realizations, analyst recognition, and index flows. On the crash side, the mechanism is the mirror image: a death-dominated rewiring documents a regime \emph{unwinding}---counterparty relationships lapsing from the filings while trailing returns still carry the regime's hot tail---so the price still reflects a configuration that the firm's own disclosures say is dissolving. The crash is the catch-down, and it is forecastable for the same reason the alpha is: the market reads prices faster than it reads filings. This connects our measure to the opacity--crash literature \citep{hutton2009opacity} with the sign sharpened: it is not general opacity but \emph{documented disintegration under a hot tail} that flags the crash.

The shared mechanism makes parallel predictions that discipline the empirical design: both effects should be concentrated where tail heat is high; both should be conditional on the network state the market is slow to read; and both should decay as configuration information becomes cheap to observe. Section~\ref{sec:theory} formalizes this; Section~\ref{sec:design} pre-specifies the tests.

\subsection{Contributions}

\begin{enumerate}[label=\textbf{C\arabic*.},leftmargin=2.6em]
\item \textbf{A conditional resolution of the lottery-tail puzzle, with two deliverables.} We provide, to our knowledge, the first observable decomposition of upper-tail heaviness into a lottery component and a structural component, using disclosure-measured network state as the discriminator---turning the MAX anomaly from a screen into a conditioning variable. The decomposition yields one deliverable per side of the title: a \emph{crash-risk flag} (tail heat with network death) that already meets a risk-monitoring evidential standard on the pilot record, and an \emph{alpha signal} (tail heat with a forming network) that is held to the stricter confirmatory standard before any capital-allocation claim is made.
\item \textbf{Measurement.} The discriminator inherits an auditable measurement pipeline: span-grounded edges, an exact birth/death/drift decomposition of rewiring, point-in-time vintage discipline, and an archived anti-hindsight protocol (frozen prompts, hashed, no theme-specific extraction). Two audits have been executed---a recall audit with sector-complete targets (Section~\ref{sec:design}) and the Stage-B0 sparsity diagnosis---while the cutoff-matched extractor audit remains pending.
\item \textbf{A pre-specified confirmatory design.} Winner-cluster threshold calibration is performed walk-forward---winners are identified inside training windows only---and the double-condition flag is benchmarked against the screens an investor already has (MAX, momentum, volatility), with the power analysis archived before estimation.
\end{enumerate}

\section{Related Work}\label{sec:related}

\paragraph{Lottery stocks and skewness preference.} \citet{bali2011max} document that stocks with the highest recent maximum daily returns underperform (the MAX anomaly); \citet{barberis2008lotteries} supply the behavioral pricing mechanism (probability weighting over skewed payoffs). Our contribution is not to dispute the discount but to condition it: we ask which tails deserve it.

\paragraph{Crash risk.} \citet{chen2001crash} show crashes are forecastable from trading-based variables; \citet{hutton2009opacity} show crash risk concentrates in firms with opaque financial reporting---bad news is withheld until it arrives at once. Our crash-side claim sharpens the disclosure channel: the danger sign is not the absence of disclosure but its \emph{content in motion}---counterparty relationships lapsing from the filings while the trailing return tail is still hot. Where the opacity literature reads how much firms say, we read what their stated economic configuration is doing.

\paragraph{Tail risk and asset prices.} \citet{kelly2014tail} extract a common tail-risk factor from the return cross-section; conditional EVT models the time-varying tail directly \citep{mcneilfrey2000,chavez2005,davison1990}, with the peaks-over-threshold foundation of \citet{balkema1974} and \citet{pickands1975}. We use standard EVT machinery (Hill and GPD tail indices on rolling windows) as the \emph{signal} layer rather than the risk-measurement layer.

\paragraph{Text, networks, and slow diffusion.} \citet{hoberg2016tnic} build firm networks from 10-K product-description similarity; \citet{cohen2008links} show customer-link return predictability; \citet{cohen2020lazy} show filing \emph{changes} predict returns because text information diffuses slowly---the friction our mechanism requires. On the prediction side, \citet{ke2020text} extract return-relevant sentiment from news with a supervised topic model, and \citet{lopezlira2023} show a general-purpose LLM's headline readings predict returns; both target the conditional \emph{mean} with text-as-sentiment. Our use of LLMs is different in kind: the model is a \emph{measurement instrument} for network structure---\citet{yang2026lem} provide the pipeline: directed, relationship-typed, span-grounded economic graphs from disclosures---and the predictive object is the tail, not the mean. The registered companion design (first author) develops the conditional-EVT framework and executed its Stage-A measurement; the present paper consumes that layer and supplies the pricing question.

\section{Theory: Two Kinds of Heavy Tails}\label{sec:theory}

\subsection{Setup}

Fix the filtered space $(\Omega,\F,\{\F_t\},\Prob)$. Firm $i$'s daily market-excess log return $r_{i,t}$ has conditional upper tail in the maximum domain of attraction with index $\xi^{+}_{i,t}$; the rolling estimate $\hat\xi^{+}_{i,t}$ (Hill estimator on the top 5\% of a trailing two-year window; Section~\ref{sec:measurement}) is the paper's \emph{tail heat} signal. The firm's economic configuration is the disclosure-measured graph state of the registered companion: adjacency $A_t$, and between consecutive filing vintages the exact rewiring decomposition
\begin{equation}
B_{i,t}+D_{i,t}+C_{i,t}\;=\;\sum_{e\in\mathcal{P}_i}\bigl|w_{e,t}-w_{e,t-1}\bigr|,
\label{eq:bdc}
\end{equation}
where $B$ is edge-birth mass and $D$ edge-death mass (both non-negative by construction: sums of born and lapsed edge weights respectively), $C$ is the absolute drift of continuing edges, and $\mathcal{P}_i$ collects directed pairs involving $i$ in either direction. Total rewiring is a magnitude, never a signed net: a birth cannot be cancelled by a continuing-edge decline.

\subsection{A two-component tail}

\begin{definition}[Tail decomposition]\label{def:twokinds}
Firm $i$'s upper-tail exceedance intensity at $t$ admits the decomposition
\begin{equation}
\lambda^{+}_{i,t} \;=\; \underbrace{\lambda^{L}_{i,t}}_{\text{lottery: transient jump risk}} \;+\; \underbrace{\lambda^{S}_{i,t}}_{\text{structural: regime repricing}},
\end{equation}
where the lottery component is serially transient (jump arrivals with no persistence in conditional mean) and the structural component is generated by repricing of a persistent configuration change: $\lambda^{S}_{i,t}>0$ only while the firm's economic network is materially recomposing and the recomposition is not yet fully priced.
\end{definition}

The components are not separately observable from returns: both raise $\hat\xi^{+}$, both raise MAX-type statistics, and at any $t$ the return history is consistent with either mixture. This unidentifiability from prices alone is, we contend, precisely why the market prices the pooled object---and why a text-measured discriminator can carry alpha.

\begin{assumption}[Pricing of the pooled tail]\label{ass:pricing}
Skewness-preferring investors price observed tail heat $\hat\xi^{+}_{i,t}$ with a discount $\delta(\hat\xi^{+}_{i,t})>0$ increasing in tail heat, applied to the pooled intensity without separating $\lambda^{L}$ from $\lambda^{S}$ \citep{barberis2008lotteries,bali2011max}.
\end{assumption}

\begin{assumption}[Slow diffusion of configuration]\label{ass:diffusion}
The network state $(B_{i,t}, D_{i,t}, A_{i,t})$ is documented in filings at vintage $t$ but is incorporated into prices with delay, as with other filing-text information \citep{cohen2020lazy}.
\end{assumption}

\begin{proposition}[Conditional alpha of tail heat]\label{prop:alpha}
Under Assumptions~\ref{ass:pricing}--\ref{ass:diffusion}, expected forward abnormal returns conditional on tail heat satisfy:
\begin{enumerate}[label=(\roman*)]
\item \textbf{(Lottery discount.)} Unconditionally, $\E[R^{\abn}_{i,t+h}\mid \hat\xi^{+}_{i,t}\ \mathrm{high}] \le \E[R^{\abn}_{i,t+h}]$: pooled tail heat earns at most the discount, since the pool is dominated by $\lambda^{L}$.
\item \textbf{(Structural premium.)} Conditional on a forming configuration (birth-dominated rewiring, intact network), high tail heat earns a non-negative and transiently positive premium: the discount $\delta$ is applied to a tail whose $\lambda^{S}$ component implies persistent repricing, and the mispricing corrects as the configuration diffuses.
\item \textbf{(Decay side.)} Conditional on a disintegrating configuration (death-dominated rewiring), high tail heat predicts \emph{negative} forward returns: the hot tail is the unwinding side of a regime, and $D$ marks the unwind before prices complete it.
\end{enumerate}
\end{proposition}

\begin{remark}
Proposition~\ref{prop:alpha} is a pricing argument, not a theorem about the data-generating process; its role is to fix signs ex ante for the empirical design. Its content is falsifiable: (i) predicts flat-to-negative unconditional tail-heat quintiles; (iii) predicts a negative $\hat\xi^{+}\times D$ interaction; (ii) predicts that the surviving cell---hot tail, birth-dominated or stable network---outperforms both the lottery cell and the unconditional average.
\end{remark}

\subsection{Which heavy tails produce essential alpha}\label{sec:essential}

The paper's answer to its central question, stated as three necessary conditions. A heavy upper tail produces \emph{essential} (structural, ex-ante harvestable) alpha exactly when:

\begin{enumerate}[label=\textbf{E\arabic*.},leftmargin=2.6em]
\item \textbf{The tail is configuration-generated.} The extreme returns coincide with material recomposition of the firm's documented economic network---edge births into expanding counterparties, new dependency types (platform, cloud, foundry), rising neighborhood weight---rather than with news-flow jumps on an unchanged network. Observable: birth-dominated rewiring, $B_{i,t}\gg D_{i,t}$, in the vintage window preceding or overlapping the tail heat.
\item \textbf{The configuration is documented but not yet common knowledge.} The edges exist in filings (span-grounded, auditable) while the market's pricing of the tail still matches the lottery template---i.e., the stock trades with the MAX-style discount despite E1. Observable: high $\hat\xi^{+}$ \emph{and} high MAX statistics with no offsetting valuation premium; the alpha window closes as recognition arrives, so the signal must decay with configuration age.
\item \textbf{The network is not on its decay side.} Death mass is low: the same tail-heat statistic attached to $D$-dominated rewiring is the \emph{crash} signature (NVIDIA 2018 and 2022 in our panel), and no threshold on $\hat\xi^{+}$ alone can separate the two. Observable: the double condition $\hat\xi^{+}\ge u^{*}$ and $D\le d^{*}$ (equivalently, birth share $B/(B+D)$ high).
\end{enumerate}

In one sentence: \emph{the heavy tail that pays is the one the market is discounting as a lottery while the filings document a forming regime---heat plus births, without deaths.} The crash-side answer is the mirror, and it needs no pricing assumption beyond slow diffusion: \emph{the heavy tail that crashes is the one attached to a disintegrating configuration---heat plus deaths---because the price still reflects a regime the firm's own filings say is dissolving.} Everything else---heat without configuration change---is noise to be discounted, and the data below cannot tell any of the three apart without the network.

\subsection{Hypotheses}

With $Z^{\mathrm{ctrl}}$ = (MAX, momentum, realized volatility) and firm-clustered inference throughout:

\begin{hypothesis}[Lottery baseline]\label{hyp:a1}
Unconditional tail heat carries no positive premium: forward abnormal returns are non-increasing across $\hat\xi^{+}$ quintiles controlling for $Z^{\mathrm{ctrl}}$, and tail asymmetry $\hat\xi^{+}-\hat\xi^{-}$ carries a non-positive premium.
\end{hypothesis}

\begin{hypothesis}[Separation, danger side]\label{hyp:a2}
The interaction $\hat\xi^{+}\times D$ predicts forward abnormal returns negatively, controlling for $Z^{\mathrm{ctrl}}$ and main effects. This is confirmatory endpoint P1 (Section~\ref{sec:design}). The complementary prediction---$\hat\xi^{+}\times B$ non-negative---is a \emph{secondary} endpoint: the pilot cannot sign it precisely, and the design does not pretend otherwise.
\end{hypothesis}

\begin{hypothesis}[Winner signature, alpha side]\label{hyp:a3}
Historical extreme winners disproportionately exhibit, before their runs, the joint signature of Section~\ref{sec:essential}: threshold-crossing tail heat with birth-dominated rewiring. Calibrated walk-forward on unified labels, the double-condition flag attains higher out-of-sample precision for forward extreme upside than (a) the tail threshold alone and (b) MAX/momentum/volatility screens. This is confirmatory endpoint P2, tested in a gatekeeping sequence after P1 (Section~\ref{sec:design}); it is the paper's economically central claim, and P1 is its powered statistical gate---the mean-regression interaction is where the pilot supplies an effect-size anchor and where power is calculable, while P2 carries the extreme-upside content the title promises.
\end{hypothesis}

\section{Measurement}\label{sec:measurement}

\subsection{Tail heat}

For each firm-month, $\hat\xi^{+}_{i,t}$ is the Hill estimator on the top order statistics of positive daily market-excess log returns in a trailing $W=504$ trading-day window, with the exact rule $k=\max(10,\lceil 0.05\,W\rceil)=26$; $\hat\xi^{-}$ symmetrically on losses; asymmetry is the difference. Two noise sources are acknowledged up front: $k$ is small, and monthly estimates from overlapping windows are heavily autocorrelated---this is why identification runs at the firm-vintage level (Section~\ref{sec:pilot}) and why inference clusters by firm. Pre-specified sensitivity: $W\in\{252,756\}$ and GPD maximum-likelihood tail indices at the same threshold. Because a trailing window mechanically \emph{contains} the early run, all event-time statements are made relative to first threshold crossing, and the confirmatory design measures forward returns strictly after signal formation.

\subsection{Network state}

The graph layer is inherited from the registered companion design and its executed Stage~A: 10-K filings parsed per vintage (Items 1, 1A, 7); a fixed, direction-neutral, theme-free extraction prompt (frozen and hashed before any return data were joined; SHA-256 archived); edges restricted to a fixed panel with span-grounded quotes; deterministic per-filing merge; rewiring components computed per \eqref{eq:bdc} over both dependency directions.

\paragraph{Deviations log (authoritative archive: companion v0.5).} The companion exists in two states: the registered protocol (v0.4, pre-execution) and the executed version (v0.5), whose Sections 6.1 and 7.1 document the Stage-A run \emph{with} its disclosed deviations; v0.5 is the authoritative archive for every number cited here. Table~\ref{tab:devlog} reconciles what was registered against what was executed.

\begin{table}[ht]
\centering\small
\begin{tabular}{@{}lll@{}}
\toprule
 & Registered (v0.4) & Executed Stage A (v0.5, used here) \\
\midrule
Universe & 42 Nasdaq-100 firms & 24-firm US tech convenience panel \\
Vintages & FY2016--FY2023 & FY2015--FY2024 (224 filings) \\
Extractor & multi-agent fusion (LEM) & single-pass DeepSeek V3, det.\ merge \\
Edges & --- (pending) & 858 span-grounded, 144 firm pairs \\
Cutoff audit (P3) & required, decision rule fixed & \emph{not executed}; results descriptive \\
Thresholds & walk-forward & pooled full-sample (descriptive) \\
\bottomrule
\end{tabular}
\caption{Deviations of the executed Stage A from the registered protocol, all disclosed in companion v0.5. Every pilot statistic in this paper is descriptive in part because of these deviations.}
\label{tab:devlog}
\end{table}

The companion's executed Stage-A intensity results cited in Section~\ref{sec:pilot-h2} are, with numbers: in the joint intensity model for unsigned exceedances, the graph rewiring coefficient at the 6-month horizon is $+0.66$ per SD (SE $0.20$) alongside document-similarity change at $+0.58$ (SE $0.17$); at 12 months the graph coefficient is $+0.38$ (SE $0.20$) while the text coefficient is $+0.08$ (SE $0.24$); the two measures correlate at $0.19$. These are descriptive, from the same panel and deviations as above. The anti-hindsight protocol---vintage-frozen normalizations, walk-forward ordering, cutoff-matched extractor audit with a pre-committed decision rule---is stated in full in the archived companion; Section~\ref{sec:design}(f) re-sizes, rather than merely inherits, the audit plan for the confirmatory universe.

\subsection{Returns and labels}

Forward $h$-month abnormal return $R^{\abn}_{i,t,h}=R_{i,t,h}-\beta_{i,t}R_{m,t,h}$ with trailing 24-month beta; horizons $h\in\{3,6,12\}$ months with $h=6$ primary; extreme-upside labels for Hypothesis~\ref{hyp:a3} take the threshold from the companion's convention (training-window $Q_{0.90}$ of $|R^{\abn}|$) but the event itself is strictly positive: $R^{\abn} > +u$, never an absolute-value exceedance (Section~\ref{sec:design}(d)).

\section{Pilot Evidence (Descriptive)}\label{sec:pilot}

All results in this section are descriptive: pooled (not walk-forward) standardizations, a 24-firm convenience panel, overlapping horizons, no multiplicity-adjusted claims. Their role is to fix signs and calibrate the confirmatory design, not to confirm it.

\paragraph{Sample accounting.} The tail panel comprises 3{,}336 firm-months; 3{,}192 have a valid forward return and controls (the Hypothesis-\ref{hyp:a1} sample); of these, 757 lack a lagged graph vintage---months before a firm's second extracted filing, a purely calendar-mechanical rule with no dependence on outcomes---leaving 2{,}435 firm-months (200 firm-vintages, 24 firms) for the interaction analysis. Because the graph state varies only at annual filing vintages, \emph{the honest count of independent identifying units is closer to 200 than to 2{,}435}, and every headline statistic below is therefore reported at both levels.

\subsection{The lottery baseline, at a boundary condition (H1)}\label{sec:pilot-h1}

Across 3{,}192 firm-months (2014--2025), forward 6-month abnormal returns by $\hat\xi^{+}$ quintile are flat: $2.7$, $2.1$, $2.7$, $2.5$, $2.3\%$. In the joint regression with controls, tail heat is dead ($b=+0.005$ per SD, SE $0.006$) while MAX is negative with marginal significance ($b=-0.007$, SE $0.004$, $t\approx-1.8$)---\emph{consistent with} the lottery discount reaching even into a mega-cap technology panel, though we do not call this a replication: the MAX anomaly is strongest among small, illiquid stocks, and a 24-firm survivor-tilted mega-cap panel is a boundary condition for it, not a test of it. Raw quintiles of tail \emph{asymmetry} slope toward underperformance (the two most right-skewed quintiles earn $0.6$--$1.4\%$ versus $3.4$--$3.7\%$ for the other three), but the effect attenuates with controls: $b=-0.004$ per SD, 95\% CI $[-0.019,+0.011]$. The modest conclusion that survives: nothing in this panel rewards unconditional tail heat, and a screen built on it alone has no support here.

\subsection{The winner in event time}

\begin{figure}[ht]
\centering
\includegraphics[width=0.9\linewidth]{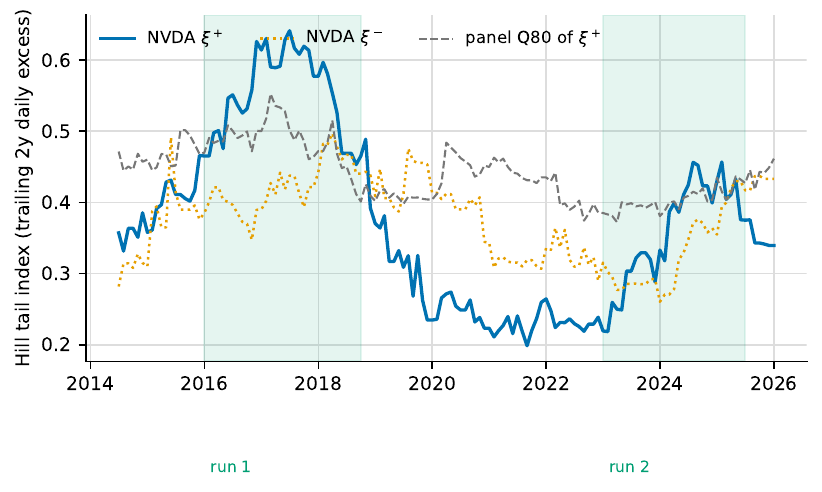}
\caption{NVIDIA rolling tail indices (Hill, trailing 504 days of daily market-excess returns) against the panel's rolling Q80 of $\hat\xi^{+}$. Shaded spans mark the two runs. Descriptive.}
\label{fig:trajectory}
\end{figure}

Figure~\ref{fig:trajectory} shows NVIDIA's $\hat\xi^{+}$ trajectory against the panel's rolling 80th percentile, with exact crossing months computed from the panel: the index first crosses the threshold in \textbf{February 2016}---at the onset of the 2016--18 run and ahead of the bulk of its return---remains elevated through the run, collapses through the 2019--2022 trough, and rises through the 2023--25 run but crosses only in \textbf{May 2024}, well after that run began. The record is therefore one timing lead and one timing lag, and we report both: a winners-calibrated threshold would have flagged NVIDIA usefully once and late once. The loss tail $\hat\xi^{-}$ shows no comparable pattern in either episode. And the flat cross-section of Hypothesis~\ref{hyp:a1} still bites: the same flag, unconditioned, also buys every lottery in the panel.

\subsection{The network separates the tails (H2, in pilot)}\label{sec:pilot-h2}

Table~\ref{tab:interaction} reports the pilot's central regression at both levels of aggregation: the monthly panel (2{,}435 firm-months) and the collapsed firm-vintage panel (200 firm-vintages)---the latter being the honest unit of identification, since the graph state varies only at filings. Standardization rule: all variables are z-scored on the pooled pilot sample, and each interaction is the \emph{product of standardized main effects} (not a re-standardized product), so interaction coefficients read as the return effect of a one-SD tail-heat move at a one-SD death-mass state.

\begin{table}[ht]
\centering\small
\begin{tabular}{@{}lrrrcrrr@{}}
\toprule
 & \multicolumn{3}{c}{Monthly ($N=2{,}435$, $R^2=0.028$)} & &
   \multicolumn{3}{c}{Firm-vintage ($N=200$, $R^2=0.194$)} \\
\cmidrule{2-4}\cmidrule{6-8}
 & $b$ & SE & $t$ & & $b$ & SE & $t$ \\
\midrule
$\hat\xi^{+}$          & $+0.010$ & $0.008$ & $+1.3$ & & $+0.013$ & $0.007$ & $+1.9$ \\
birth mass $B$          & $+0.015$ & $0.011$ & $+1.4$ & & $+0.009$ & $0.007$ & $+1.2$ \\
death mass $D$          & $-0.009$ & $0.009$ & $-1.0$ & & $-0.006$ & $0.007$ & $-0.9$ \\
$\hat\xi^{+}\times D$   & $\mathbf{-0.025}$ & $0.009$ & $\mathbf{-2.9}$ & & $\mathbf{-0.028}$ & $0.007$ & $\mathbf{-3.9}$ \\
$\hat\xi^{+}\times B$   & $-0.000$ & $0.007$ & $-0.1$ & & $+0.005$ & $0.008$ & $+0.7$ \\
momentum                & $-0.009$ & $0.012$ & $-0.7$ & & --- & --- & --- \\
volatility              & $+0.010$ & $0.010$ & $+1.0$ & & --- & --- & --- \\
MAX                     & $-0.010$ & $0.005$ & $-2.0$ & & $+0.040$ & $0.035$ & $+1.2$ \\
\bottomrule
\end{tabular}
\caption{Forward 6M abnormal return on tail heat $\times$ rewiring components; firm-clustered SEs; both aggregation levels; momentum and volatility included in the collapsed model (coefficients omitted for space; MAX changes sign on collapsing because vintage-averaged MAX is a different object than monthly MAX---flagged, not interpreted). Descriptive.}
\label{tab:interaction}
\end{table}

\paragraph{Robustness of the one number that matters.} Because 24 firm clusters make asymptotic clustered $t$-statistics unreliable, we report the battery a skeptic would demand. Wild-cluster bootstrap (Rademacher weights, null imposed) on the collapsed regression: two-sided $p=0.039$---significant, but visibly weaker than the raw $t=-3.9$ suggests, and this is the number we quote. Two-way (firm and month) clustering on the monthly panel: $t=-3.0$, essentially unchanged. Leave-one-firm-out: the interaction ranges from $-0.029$ to $-0.015$; the largest attenuation comes from dropping NVIDIA (coefficient roughly halves, sign intact)---the effect is not one firm, but one firm carries a disproportionate share, as befits a 24-firm panel. Support: 200 firm-vintages, of which 73\% have $D>0$ (median $D=0.6$, maximum $7.0$) and 84\% have $B>0$; the interaction is identified off genuinely distributed variation, not a handful of events, though the deep-death tail ($D>2$) is thin.

\begin{figure}[ht]
\centering
\includegraphics[width=0.68\linewidth]{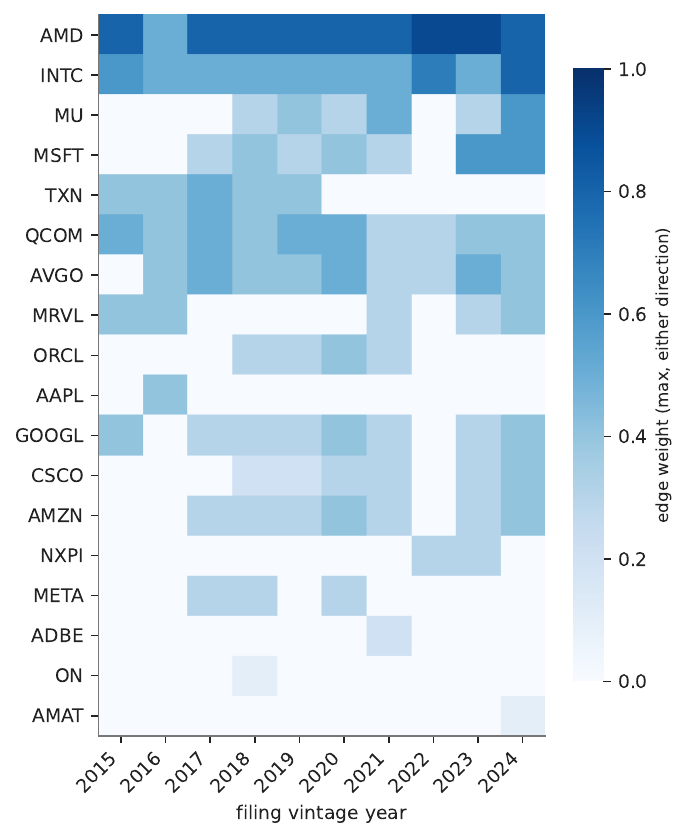}
\caption{NVIDIA's disclosure-measured neighborhood, FY2015--FY2024: maximum edge weight (either direction) per counterparty and vintage year, single-pass extraction. Darker cells indicate higher edge weight. Descriptive.}
\label{fig:neighborhood}
\end{figure}

In event terms, with exact dates: NVIDIA's FY2022 10-K (accepted 2022-03-18) was flagged death-dominated ($D=7.0$ of a total rewiring of $8.7$) while its trailing tail was still hot; the forward 6-month window (2022-03-18 to 2022-09-16) delivered a $-51\%$ abnormal return. Its 2018 crypto episode fits the same template; its 2020/2021/2024 vintages---birth-heavy rewiring under rising heat---preceded $+45/+32/+34\%$ abnormal 6-month returns. Figure~\ref{fig:neighborhood} shows the underlying measured neighborhood: the 2022 flag is visible as the simultaneous decay of several 2018-era edges, and the 2020--2024 flags as the strengthening Microsoft/Micron cluster.

\begin{figure}[ht]
\centering
\begin{minipage}{0.495\linewidth}
\centering
\includegraphics[width=\linewidth]{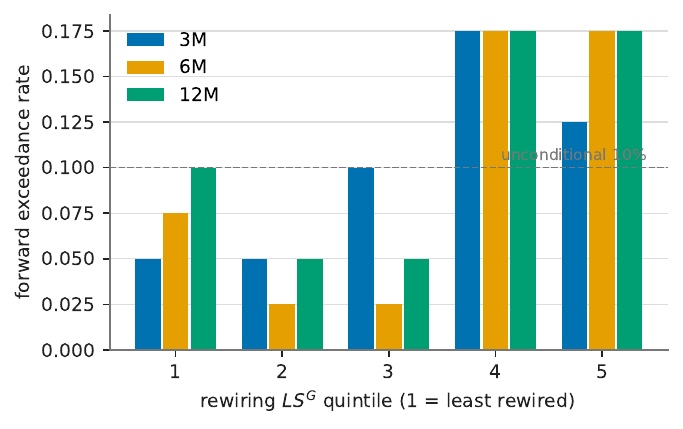}
\end{minipage}\hfill
\begin{minipage}{0.495\linewidth}
\centering
\includegraphics[width=\linewidth]{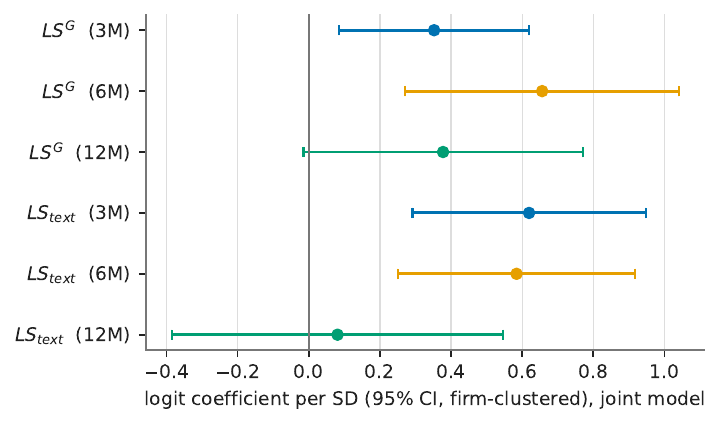}
\end{minipage}
\caption{Companion Stage-A intensity exhibits (descriptive; companion v0.5 §7.1). Left: forward unsigned exceedance rate by rewiring ($LS^G$) quintile and horizon; dashed line marks the 10\% unconditional rate. Right: joint intensity model coefficients per SD with 95\% firm-clustered intervals---rewiring ($LS^G$) and document-similarity change ($LS_{\mathrm{text}}$) in the same regression.}
\label{fig:companion}
\end{figure}

The companion's executed Stage-A intensity results (Section~\ref{sec:measurement}: graph coefficient $+0.66$ vs.\ text $+0.58$ per SD at 6M; $+0.38$ vs.\ $+0.08$ at 12M; correlation $0.19$) supply the corresponding descriptive evidence that the network state is not repackaged return or text information; Figure~\ref{fig:companion} reproduces the two companion exhibits---forward exceedance rates rising in rewiring quintiles, and the joint-model coefficients in which the graph survives at 12 months where document-similarity change dies.

\subsection{What the pilot does not show}

The $D$-interaction is one coefficient in one small panel, estimated with pooled standardization; the positive cell of Proposition~\ref{prop:alpha}(ii) is directionally present (birth-heavy heat earns no discount) but not significant; a double-sort by birth mass among heavy-tail months is noisy and sensitive to controls. The panel is winner-tilted by construction (large-cap survivors), which \emph{biases toward} finding structural tails. None of the pilot facts are confirmatory claims.

\subsection{An adverse pre-registered replication (Stage B0)}\label{sec:b0}

Before any scale-up, we ran a budget-bounded ($\approx$\$6) replication with a protocol frozen in advance (seed, sampling frame, regression, and success criterion archived before any return data were joined): 50 firms drawn uniformly at random from the S\&P~500 (a free proxy for the large-cap Russell segment), \emph{excluding all 24 pilot firms}, FY2020--FY2024 vintages, identical extraction prompt, identical collapsed regression, success defined as a negative $\hat\xi^{+}\times D$ coefficient with one-sided wild-cluster $p<0.10$.

\textbf{The replication failed under its frozen criterion}: the interaction coefficient is $-0.001$ ($t=-0.07$; one-sided wild-cluster $p=0.49$; $N=200$ firm-vintages, 50 firms). We report this verbatim, as the protocol requires. The failure mode, however, is itself a finding---and it is the one the protocol's sparsity clause anticipated: on random cross-sector firms the disclosure graph nearly vanishes. The 255 replication filings yielded 234 edges over 57 firm pairs (the coherent 24-firm technology pilot yielded 858 over 144); \textbf{83\% of replication firm-vintages have zero death mass} (median $D=0$), against 27\% in the pilot, far beyond the 60\% underpowered-by-sparsity threshold declared in the protocol. REITs, pharmaceutical firms, and utilities rarely name panel counterparties in their 10-Ks at all.

Two interpretations are observationally equivalent in this data, and we decline to choose between them: (i) the pilot effect is specific to coherent technology ecosystems (adverse to generality); (ii) the \emph{measurement} has no support outside dense disclosure neighborhoods---the discriminator cannot fail or succeed where it does not exist (adverse to the instrument's coverage, not necessarily to the effect). What the replication establishes either way is a scope condition that now bounds the paper's claims: \emph{the two-kinds-of-heavy-tails mechanism is testable only where firms document their economic counterparties densely}---coherent supply-chain and platform ecosystems---and any confirmatory universe must be constructed, and any eventual result qualified, accordingly.

The natural next check---does the interaction survive inside the replication's own dense subset?---is uninformative at replication scale, and we report it rather than lean on it: among the six dense technology firms of the random draw (24 firm-vintages) the interaction coefficient is $-0.001$ (SE $0.057$); among all 34 firm-vintages with $D>0$ it is $-0.040$ (SE $0.053$). Point estimates spanning zero to the pilot magnitude, with standard errors an order of magnitude wider than the pilot effect, decide nothing; the dense-subset test is delegated to the confirmatory stage. Two further qualifications: the S\&P~500 sampling frame is itself a proxy for the Russell large-cap segment (constituent licensing), so the scope condition is re-measured on the actual confirmatory universe by the frozen density gate of Section~\ref{sec:design}; and all replication densities are lower bounds, since some zeros reflect target-eligibility rather than absent disclosure (the recall audit of Section~\ref{sec:design} quantifies exactly this).

\begin{figure}[p]
\centering
\includegraphics[height=0.92\textheight]{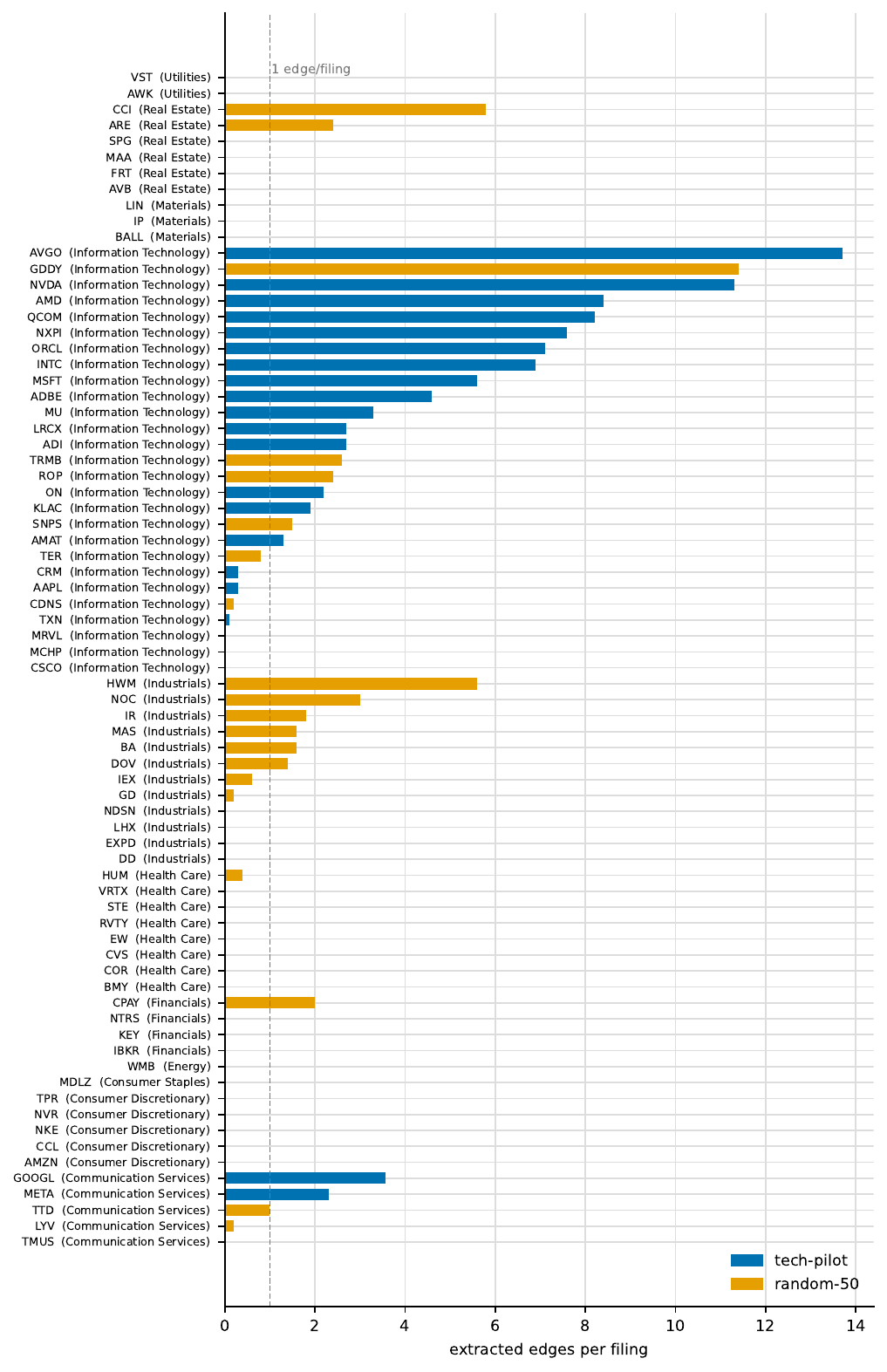}
\caption{Disclosure-graph density map across the two extracted corpora: edges per filing by firm and GICS sector (blue: 24-firm technology pilot; orange: 50-firm random draw). Dashed line marks one edge per filing. Densities are lower bounds---some zeros reflect target-panel eligibility rather than absent disclosure (Section~\ref{sec:b0}).}
\label{fig:density}
\end{figure}

Figure~\ref{fig:density} maps the scope condition firm by firm across both corpora. Three of its features discipline the design going forward. Density is a \emph{sector} property, not a pilot artifact: the random draw's own technology firms---never touched by the pilot---are dense (up to 11 edges per filing), while health care, consumer, energy, materials, and utilities are near-zero regardless of cohort. Unexpected pockets of density exist outside technology (aerospace--defense primes; telecom-infrastructure firms that name their carrier customers). And ecosystems are hub-centered rather than sector-shaped: 65\% of the random panel's edges point at mega-cap hubs rather than at other sampled firms, so a viable confirmatory universe is a hub-neighborhood completion, not a sector block.

\section{Confirmatory Design}\label{sec:design}

\textbf{Status: on hold, with the re-activation path now executed or frozen.} Per the pre-committed consequence clause of the Stage-B0 protocol (Section~\ref{sec:b0}), the confirmatory stage does not run until the replication discrepancy is understood. Since v0.3, the two diagnostics have been carried out and their consequences frozen:

\emph{(i) Recall audit --- executed, verdict binding.} Twelve zero-edge filings (three technology, three non-technology firms, two filings each) were re-extracted under a protocol frozen in advance, with firm-specific sector-complete target lists; the pre-committed rule (fewer than 50\% of filings gaining edges $\Rightarrow$ single-pass recall inadequate) triggered at 2/12. \textbf{The confirmatory stage therefore uses multi-agent extraction.} A word-boundary text cross-check decomposes the twelve: two genuine extraction-recall failures (CVS: competitors named in text, targets eligible, zero edges returned), two \emph{parser}-coverage failures (Cisco: competitor names present in the raw filing but outside the captured Item spans---adding a parser-validation prerequisite), six legitimately nameless documents (no fix exists; the scope condition covers them), and two eligibility artifacts cured by sector-complete targets (Williams: midstream-peer edges appeared). The extractor for the confirmatory stage is re-frozen as \texttt{deepseek-v4-flash} (model identifier archived per the P1 protocol).

\emph{(ii) Universe construction --- frozen.} The ecosystem-coherent universe is defined by an archived rule (three membership criteria: S\&P~1500 Information Technology and Communication Services; S\&P~1500 Aerospace \& Defense; a fixed telecom-infrastructure list), point-in-time at each June reconstitution, 10-K filers only, with \emph{sector-complete target lists} (all universe members plus the mega-cap hub list) mandated by the audit finding. A pre-registered \emph{density gate} precedes any confirmatory data assembly: a seeded 200-filing pre-test must show a firm-vintage $D>0$ share of at least 30\%; if it fails, the universe rule must be revised, re-frozen, and re-tested before any return data are joined. This gate is the single remaining condition for re-activation, and the scope qualification---claims speak only to densely disclosing ecosystems---attaches to every eventual result.

The design inherits the companion protocol's discipline (vintage-frozen statistics, walk-forward ordering, archived power analysis before estimation) and re-scopes it around a \emph{gatekept primary pair}. Everything below was frozen before any confirmatory data were assembled.

\begin{enumerate}[label=(\alph*)]
\item \textbf{Universe and data.} The frozen ecosystem-coherent universe (status note above): S\&P~1500 Information Technology and Communication Services members, S\&P~1500 Aerospace \& Defense, and the fixed telecom-infrastructure list, point-in-time at each June reconstitution, 2010--2024; 10-K filers only (foreign private issuers filing 20-F excluded); 10-K/A amendments enter as of their own acceptance dates, originals never replaced retroactively. Graph extraction covers FY2008--FY2024 ($\approx$6{,}500--7{,}000 filings), \emph{multi-agent} per the audit verdict, with sector-complete target lists and the parser-validation prerequisite; the density gate must pass first. The adversarial edge audit is \emph{re-sized}, not inherited: sample size is set by a $\pm0.07$ binomial margin at 95\% confidence ($n\approx200$ audited edges), stratified by filing year and GICS sector; the cutoff-matched extractor audit and its pre-committed decision rule (companion Assumption P3) execute \emph{before} any return data are joined.
\item \textbf{Panel construction and the common-covered-sample rule.} Monthly tail panel with $W=504$, $k=26$ (Section~\ref{sec:measurement}); graph state carried forward from the most recent vintage at or before month-end, with a staleness covariate (months since acceptance) always included; firm-months before a firm's second extracted vintage are excluded ($D$ undefined). \emph{Common-covered-sample rule:} every model, screen, and benchmark in (d)--(f) is evaluated on the identical set of firm-months possessing (i) a valid lagged graph state, (ii) a valid tail estimate ($\ge$504 days of price history), and (iii) a valid forward return---membership is determined by these data-availability conditions alone, never by outcomes, and coverage is reported per period.
\item \textbf{Walk-forward schedule (fixed windows, one calibration).} Training 2010--2017, validation 2018--2019, test 2020--2024, with \emph{no refitting}: all z-scores, winsorization bounds (1st/99th percentiles of forward returns), thresholds, and the P2 calibration pair are computed once---standardizations and thresholds on the training window, the P2 grid selection on the validation window---and then frozen for the entire test period. No test-period outcome enters any standardization, threshold, calibration, or hyperparameter choice; the P1 regression \emph{coefficient} is, by construction, estimated on test-period observations (that is what testing means), using only frozen inputs. Interactions are products of training-standardized main effects. Fixed windows are chosen over per-year refits deliberately: they remove any path by which test years could enter estimation, at the acknowledged cost of staler calibrations in late test years.
\item \textbf{Primary endpoints (gatekept pair; family-wise $\alpha=0.05$).}
  \begin{itemize}
  \item \textbf{P1 (separation; Hypothesis~\ref{hyp:a2}).} The coefficient on $z(\hat\xi^{+})\times z(D)$ in the \emph{firm-vintage collapsed} regression of winsorized forward-6M abnormal return on $\{z\hat\xi^{+}, zB, zD, z\hat\xi^{+}zB, z\hat\xi^{+}zD, z\mathrm{mom}, z\mathrm{vol}, z\mathrm{MAX}, \mathrm{staleness}\}$, estimated on the test period. \emph{Decision rule:} one-sided (negative); wild-cluster bootstrap $p<0.05$ (Rademacher, null-imposed, 9{,}999 replications, firm clusters), with the firm-cluster bootstrap confidence interval and two-way (firm, month) clustered variant reported alongside.
  \item \textbf{P2 (winner signature; Hypothesis~\ref{hyp:a3}), tested only if P1 is supported.} Precision-at-$k$ of the double-condition flag for the \emph{extreme-upside label}, with $k=20$ names per test year. \emph{Unified, positive-only labels:} both the winners used in calibration and the P2 outcome are defined identically as $R^{\abn}_{12M} > +u$ with $u$ the training window's $Q_{0.90}$ of $|R^{\abn}|$---the threshold is set on the absolute-return scale but the event is strictly positive, so large crashes can never qualify as winners in either calibration or evaluation. Benchmarks at the same $k$: tail-threshold-only, low-MAX, momentum, and low-volatility screens. \emph{Decision rule (selection-corrected):} the statistic is flag precision minus the \emph{maximum} benchmark precision, with the maximum recomputed \emph{inside every bootstrap replication}, so the comparison prices the selection over four benchmarks rather than inheriting an inflated null; each benchmark is also reported individually as secondary. \emph{Bootstrap, fully specified:} two-stage block scheme---resample the five test years with replacement, then within each drawn year resample that year's flagged and benchmark-selected names with replacement---$B=9{,}999$ replications, one-sided at 95\% (supported iff the 5th percentile of the statistic exceeds zero). With five test years this resampling is coarse, which the P2 power analysis in (g) prices explicitly.
  \end{itemize}
  The sequence preserves family-wise error at 5\%: P2 inherits P1's $\alpha$ only upon P1's success. The division of labor mirrors the title: P1 carries the paper's \emph{crash-risk} claim (the powered gate, with the pilot supplying its effect-size anchor), and P2 carries its \emph{alpha} claim (the extreme-upside content). The ordering is itself substantive: the crash side is tested first because it is the side the pilot already supports, and the alpha claim is entertained only in a world where the separation mechanism has confirmed.
\item \textbf{Calibration grid (for P2), frozen.} $u^{*}\in\{Q_{70},Q_{80},Q_{90}\}$ of training-window $\hat\xi^{+}$; network condition $\in\{D\le \mathrm{median}, D\le Q_{75}, B/(B{+}D)\ge0.5, B/(B{+}D)\ge0.75\}$---a 12-cell grid, deliberately small to bound overfitting. The pair is selected on the \emph{validation} period (not training, and never test) by precision-at-$k$; ties resolve to the more conservative cell (higher $u^{*}$, stricter network condition); the selected pair is then frozen for all test years. The overfitting risk of even this selection is disclosed: with 12 cells and one validation period, the winner's validation precision is upward-biased, which is why P2's test is against benchmarks rather than against zero.
\item \textbf{Secondary family (Benjamini--Hochberg, $q=0.10$).} $\hat\xi^{+}\times B\ge0$; horizons 3/12M; window sensitivity $W\in\{252,756\}$; GPD-MLE tail variant; the E2 decay prediction (flag strength declining in configuration age); per-tail decompositions; portfolio exhibits (double-condition long flag vs.\ danger-flag avoidance) are descriptive.
\item \textbf{Power (simulated, archived).} A simulation at the collapsed firm-vintage level ($N=1{,}000$ firms $\times$ 13 vintages, within-firm signal persistence $\rho=0.5$, firm effects, outcome dispersion at the pilot's collapsed value 0.123 and a conservative 0.20, and the exact P1 decision rule) is archived with hashed script and seed (\texttt{results/power\_interaction\_summary.md}). Results (Figure~\ref{fig:p1power}): in the base scenario, 80\% power obtains at an interaction magnitude of $|b|\approx0.004$ (simulated power $0.95$ at $-0.004$); under conservative dispersion, at $|b|\approx0.005$ ($0.74$ at $-0.004$, $0.99$ at $-0.008$). Because the scope condition shrinks and sparsifies the universe, the simulation was re-run under \emph{ecosystem-coherent scenarios}: 400 firms, zero-inflated death mass with $D>0$ shares of 60\% and 40\% (bracketing the density map's expectation). Simulated power is $1.00$ throughout the grid down to $|b|=0.010$ in every scenario---the minimum detectable effect remains below $0.010$, and the pilot anchor $-0.028$ is detected with near certainty even at the smaller, sparser scale. \emph{Pre-committed consequence:} if the density pre-test's measured $D>0$ share, fed back into this archived simulation, pushes the minimum detectable effect above the pilot anchor $|{-}0.028|$, P1 is demoted to estimation-with-interval and no confirmatory claim is made---determined before any test-period data are unblinded.

\emph{P2 power (archived, and honestly poor).} A companion simulation prices the P2 decision rule exactly as frozen (precision-at-$k{=}20$ over five test years against the per-replication maximum of four benchmarks, two-stage bootstrap): with benchmark precision near 10\%, simulated power is $0.07$ for a true precision gain of $+0.10$, $0.37$ at $+0.15$, $0.68$ at $+0.20$, and $0.89$ at $+0.25$. \textbf{P2 detects only very large improvements} (80\% power near a $+0.22$ gain), a direct consequence of 100 flag slots and five resampleable years. \emph{Pre-committed consequences:} $k$ and the year count stay frozen (no post hoc tuning toward significance); and a non-rejecting P2 with a positive point improvement is reported as \emph{underpowered-inconclusive}, never as evidence against the winner signature---P2's null carries no evidential weight against Hypothesis~\ref{hyp:a3}, only its rejection carries weight for it.
\end{enumerate}

\begin{figure}[ht]
\centering
\includegraphics[width=0.62\linewidth]{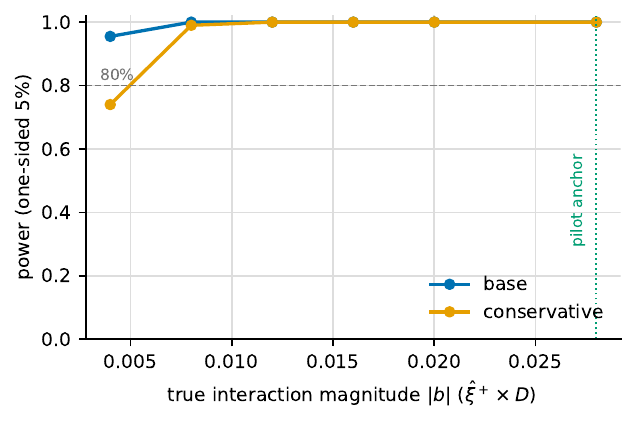}
\caption{Archived power simulation for primary endpoint P1 under the frozen decision rule, by true interaction magnitude. Dashed line marks 80\% power; dotted line marks the pilot anchor $|b|=0.028$.}
\label{fig:p1power}
\end{figure}

\section{Discussion}\label{sec:discussion}

\paragraph{Relation to the MAX anomaly.} The design does not challenge \citet{bali2011max}; it completes the anomaly's conditional structure. If Hypothesis~\ref{hyp:a2} confirms, the MAX discount is revealed as a pooled price on two objects, correctly applied to lottery tails and misapplied to structural ones---and the misapplication is detectable ex ante only through text, which is why return-based screens cannot arbitrage it away quickly (Assumption~\ref{ass:diffusion}).

\paragraph{Practical use before confirmation: the danger flag as a risk-monitoring input.} The two sides of the sign pattern do not face the same evidential bar in practice. An \emph{alpha} signal must clear confirmation before capital follows it; a \emph{risk} signal earns a place in a monitoring pool on weaker evidence, because its failure modes are asymmetric---a false danger flag costs forgone exposure, while a missed one costs a drawdown. The death-side interaction already meets the risk-pool standard on the pilot record: it is the configuration that preceded NVIDIA's 2018 and 2022 drawdowns, it survives the pilot's full inference battery (wild-cluster $p=0.04$ at the firm-vintage level), and it is cheap to compute from filings as they arrive. A position whose trailing upper tail is hot \emph{and} whose disclosure network is shedding edges warrants review regardless of what the confirmatory stage eventually concludes about the alpha side---subject to the same scope condition as every claim in this paper: the flag is defined only where disclosure is dense enough for $D$ to exist.

\paragraph{Why the network, and not fundamentals, is the discriminator.} Configuration change is visible in the graph \emph{before} it is visible in realized fundamentals: NVIDIA's data-center edges were documented years before data-center revenue dominated its mix. Fundamentals confirm regimes; disclosures document their formation. This is the ``channels before comovement'' logic of the companion design, applied to the pricing question.

\paragraph{Limitations.} Trailing tail estimation is slow and partially endogenous to the run; monthly granularity blurs flag timing; the pilot panel is small, technology-concentrated, and survivor-tilted; the graph's panel-interior restriction attenuates death detection for counterparties outside the universe; extraction was single-pass in Stage~A (the multi-agent and cutoff-audit upgrades are costed and pending); and Proposition~\ref{prop:alpha} takes the pricing mechanism from the literature rather than deriving it. The confirmatory stage addresses the first four; the last is a modeling choice made explicit.

\section{Conclusion}\label{sec:conclusion}

Heavy tails are not one thing. The market prices tail heat as if it were---discounting it wholesale as lottery exposure---and for most stocks, most of the time, that pooled price is right. But the tails of historical extreme winners were not lottery tails: they were the statistical shadows of economic reconfigurations that were documented, span by span, in the companies' own filings while the discount was still being applied. And the tails that preceded the great drawdowns were not lottery tails either: they were regimes visibly unwinding in those same filings while prices still carried the boom.

The paper's two title objects stand on deliberately different evidential footing. The \emph{crash-risk} side is the one the pilot already supports---tail heat attached to network death predicts the crash side, robustly enough to earn a place in a risk-monitoring pool today---while the \emph{alpha} side is directionally consistent but statistically unresolved, its winner narrative resting on NVIDIA's episodes rather than panel-wide evidence, and it is held behind the confirmatory gate. A pre-registered replication then showed the discriminator exists only where disclosure is dense, and the confirmatory design now carries that scope condition explicitly. If the design confirms, the payoff is symmetric: the crash flag hardens from a monitoring input into a validated predictor, and the essential alpha---the premium for reading the filings before the category error corrects---becomes harvestable. Either way, the underlying skill the paper prices is the same: knowing which kind of heavy tail one is holding, in the ecosystems where the filings say enough to tell.


\begin{thebibliography}{99}

\bibitem[Bali et~al.(2011)]{bali2011max}
Bali, T.~G., N.~Cakici, and R.~F. Whitelaw (2011). Maxing out: Stocks as lotteries and the cross-section of expected returns. \emph{Journal of Financial Economics} 99(2), 427--446.

\bibitem[Balkema and de Haan(1974)]{balkema1974}
Balkema, A.~A., and L.~de Haan (1974). Residual life time at great age. \emph{Annals of Probability} 2(5), 792--804.

\bibitem[Barberis and Huang(2008)]{barberis2008lotteries}
Barberis, N., and M.~Huang (2008). Stocks as lotteries: The implications of probability weighting for security prices. \emph{American Economic Review} 98(5), 2066--2100.

\bibitem[Chavez-Demoulin and Davison(2005)]{chavez2005}
Chavez-Demoulin, V., and A.~C. Davison (2005). Generalized additive modelling of sample extremes. \emph{Journal of the Royal Statistical Society: Series C} 54(1), 207--222.

\bibitem[Chen et~al.(2001)]{chen2001crash}
Chen, J., H.~Hong, and J.~C. Stein (2001). Forecasting crashes: Trading volume, past returns, and conditional skewness in stock prices. \emph{Journal of Financial Economics} 61(3), 345--381.

\bibitem[Cohen and Frazzini(2008)]{cohen2008links}
Cohen, L., and A.~Frazzini (2008). Economic links and predictable returns. \emph{Journal of Finance} 63(4), 1977--2011.

\bibitem[Cohen et~al.(2020)]{cohen2020lazy}
Cohen, L., C.~Malloy, and Q.~Nguyen (2020). Lazy prices. \emph{Journal of Finance} 75(3), 1371--1415.

\bibitem[Davison and Smith(1990)]{davison1990}
Davison, A.~C., and R.~L. Smith (1990). Models for exceedances over high thresholds. \emph{Journal of the Royal Statistical Society: Series B} 52(3), 393--442.

\bibitem[Hoberg and Phillips(2016)]{hoberg2016tnic}
Hoberg, G., and G.~Phillips (2016). Text-based network industries and endogenous product differentiation. \emph{Journal of Political Economy} 124(5), 1423--1465.

\bibitem[Hutton et~al.(2009)]{hutton2009opacity}
Hutton, A.~P., A.~J. Marcus, and H.~Tehranian (2009). Opaque financial reports, $R^2$, and crash risk. \emph{Journal of Financial Economics} 94(1), 67--86.

\bibitem[Ke et~al.(2020)]{ke2020text}
Ke, Z.~T., B.~T. Kelly, and D.~Xiu (2020). Predicting returns with text data. Working paper, NBER No.~26186.

\bibitem[Kelly and Jiang(2014)]{kelly2014tail}
Kelly, B., and H.~Jiang (2014). Tail risk and asset prices. \emph{Review of Financial Studies} 27(10), 2841--2871.

\bibitem[Lopez-Lira and Tang(2023)]{lopezlira2023}
Lopez-Lira, A., and Y.~Tang (2023). Can ChatGPT forecast stock price movements? Return predictability and large language models. Working paper, arXiv:2304.07619.

\bibitem[McNeil and Frey(2000)]{mcneilfrey2000}
McNeil, A.~J., and R.~Frey (2000). Estimation of tail-related risk measures for heteroscedastic financial time series: An extreme value approach. \emph{Journal of Empirical Finance} 7(3--4), 271--300.

\bibitem[Pickands(1975)]{pickands1975}
Pickands, J. (1975). Statistical inference using extreme order statistics. \emph{Annals of Statistics} 3(1), 119--131.

\bibitem[Yang and Zhang(2026)]{yang2026lem}
Yang, F., and L.~Zhang (2026). LLM latent edge measurement: Point-in-time economic graphs for quantitative investing from corporate disclosures. Working paper, arXiv:2607.15640.

\end{thebibliography}
\end{document}